\documentstyle[12pt]{article}

\begin{document}
\setcounter{page}{1}

\pagestyle{plain} \vspace{1cm}
\begin{center}
\Large{\bf Some Aspects of Minimal Length Quantum Mechanics}\\
\small
\vspace{1cm} {\bf Kourosh Nozari} \quad and \quad {\bf Tahereh Azizi}\\
\vspace{0.5cm} {\it Department of Physics,
Faculty of Basic Sciences,\\
University of Mazandaran,\\
P. O. Box 47416-1467, Babolsar, IRAN} \\
e-mail: knozari@umz.ac.ir

\end{center}
\vspace{1.5cm}

\begin{abstract}
String theory, quantum geometry, loop quantum gravity and black hole
physics all indicate the existence of a minimal observable length on
the order of Planck length. This feature leads to a modification of
Heisenberg uncertainty principle. Such a modified Heisenberg
uncertainty principle is referred as gravitational uncertainty
principle(GUP) in literatures. This proposal has some novel
implications on various domains of theoretical physics. Here, we
study some consequences of GUP in the spirit of Quantum mechanics.
We consider two problem: a particle in an one-dimensional box and
momentum space wave function for a "free particle". In each case we
will solve corresponding perturbational equations and
compare the results with ordinary solutions.

\end{abstract}
\newpage
\section{Introduction}
The problem of reconciling Quantum Mechanics with General Relativity
is one of the task of modern theoretical physics which, until now,
has not yet found a consistent and satisfactory solution. Although a
full description of quantum gravity is not yet available, there are
some general features that seem to go hand in hand with all
promising candidates for such a theory where one of them is the
existence of a minimal length scale. In other words, one of the most
interesting consequences of unification of gravity and quantum
mechanics is that in resulting quantum gravity there exists a
minimal observable distance on the order of the Planck length, $
l_{P} =\sqrt{\frac{G\hbar}{c^3}}\sim 10^{-33}cm$, where G is the
Newton constant. The existence of such a fundamental length is a
dynamical phenomenon due to the fact that, at Planck scale, there
are fluctuations of the background metric, i.e. a limit of the order
of Planck length appears when quantum fluctuations of the
gravitational field are taken into account. In the language of
string theory one can say that a string cannot probe distances
smaller than its length. The existence of such minimal observable
length which is motivated from string theory[1-7],loop quantum
gravity[8], quantum geometry[9] and black holes physics[10], leads
to a generalization of Heisenberg uncertainty principle to
incorporate gravitational induced uncertainty from very beginning.
This feature constitutes a part of the motivation to study the
effects of this modified algebra on various observables.
Consequences of such a gravitational uncertainty principle (GUP),
have been studied extensively[11-20]. The generalization of
Schr\"{o}dinger equation has been considered by Hossenfelder {\it et
al}[21]. Also generalized Schr\"{o}dinger equation and Hydrogen
spectrum in the framework of GUP have been considered by Brau[22]
and Akhoury and Yao[23]. Momentum space representation of quantum
states has been considered by Kempf {\it et al}[24]. Here we proceed
some more step in this direction. We consider some well-known topics
in usual quantum mechanics and re-examine them in the framework of
GUP. The problems of a particle in an one-dimensional box and
momentum space wave function(in a different view relative to Kempf
{\it et al} point of view) are considered  and their generalization
in GUP are discussed. \\

\section{Minimal Length and GUP}
The emergence of a minimal observable distance yields to the
generalized uncertainty principle
\begin{equation}
\label{math:2.1} \Delta x\geq \frac{\hbar}{\Delta p} + \alpha
l_{P}^2\frac{\Delta p}{\hbar},
\end{equation}
where $\alpha$ is GUP parameter which can be determined from
fundamental theory(maybe string theory!)[6]. At energy much below
the Planck mass, $ m_{P} =\sqrt{\frac{\hbar c}{G}}\sim
10^{19}GeV/c^2$, the extra term in right hand side of equation (1)
is irrelevant and the Heisenberg relation is recovered, while, as we
approach the Planck energy, this term becomes relevant and, as has
been said, it is related to the minimal observable length. A simple
calculation gives $(\Delta x)_{min} = 2l_{P}\sqrt{\alpha}$ for this
minimum length scale. Now the generalized commutation relation
becomes,
\begin{equation}
\label{math:2.2} [x,p]=i\hbar(1+\beta p^2),
\end{equation}
where $\beta$ (related to $\alpha$) is a constant restricted to the
condition: $0\leq \beta \leq 1$. The case $\beta\rightarrow 0$ gives
the usual quantum mechanics regime while $\beta\rightarrow 1$ is
extreme quantum gravity limit. It is important to note that GUP
itself has a perturbational nature and one can consider its more
generalized form[24]. Since we are dealing with dynamics, the
present form of GUP as equation (1) is more suitable as our primary
input.

\section{A particle in an One-Dimensional Box}
Consider a spinless quantum particle with mass $m$ confined to the
following one-dimensional box
\begin{equation}
\label{math:1.1} V(x)=\left\{\begin{array}{ll} 0& 0< x < a\\

\infty&{\rm elsewhere}\end{array}\right.
\end{equation}
Ordinary Schr\"{o}dinger equation for such a particle is
\begin{equation}
\frac{P_{op}^2}{2m}\psi(x)=E\psi(x),
\end{equation}
where $P_{op}$ is momentum operator. In GUP, existence of minimal
length scale leads to the following generalization of momentum
operator[22,23]
\begin{equation}
P_{op}=\frac{\hbar}{i}\frac{\partial}{\partial
x}[1+\beta(\frac{\hbar}{i}\frac{\partial}{\partial x})^2],
\end{equation}
up to first order of $\beta$. Now the Schr\"{o}dinger equation
generalizes to
\begin{equation}
\frac{1}{2m}({-\hbar}^2\frac{\partial^2\psi(x)}{\partial
x^2}+2\beta\hbar^4\frac{\partial^4\psi(x)}{\partial x^4})=E\psi(x),
\end{equation}
where can be written as
\begin{equation}
2\beta\hbar^4\frac{\partial^4\psi(x)}{\partial
x^4}-{\hbar}^2\frac{\partial^2\psi(x)}{\partial x^2}-2mE\psi(x)=0.
\end{equation}
Now we want to solve this eigenvalue problem. Existence of minimum
observable length means that one cannot have localized states. The
very notion of locality and position space representation breaks
down in GUP. However as Kempf {\it et al} have shown[24], when there
is no minimal uncertainty in momentum, one can work with the
convenient representation of the commutation relations on momentum
space wave function. Then one can define states with maximal
localization which are proper physical states. One can use them to
define a quasi-position representation. This representation has a
direct interpretation in terms of position measurements, although it
does not diagonalize position operator. Here, since we are dealing
with (1) which has no minimum uncertainty in momentum, we can use
such quasi-position states, $\psi(x)$ which are maximally localized.\\
Now the general solution of equation (7) is
$$\psi(x)=C_{1}e^{\frac{1}{2\hbar}\sqrt{\frac{1}{\beta}(1+\sqrt{1+16mE\beta
})}x}+C_{2}e^{\frac{-1}{2\hbar}\sqrt{\frac{1}{\beta}(1+\sqrt{1+16mE\beta
})}x}$$
\begin{equation}
 +C_{3}e^{\frac{i}{2\hbar}\sqrt{\frac{1}{\beta}(\sqrt{1+16mE\beta
}-1)}x}
+C_{4}e^{\frac{-i}{2\hbar}\sqrt{\frac{1}{\beta}(\sqrt{1+16mE\beta
}-1)}x},
\end{equation}
where $C_{i}$, $i=1,2,3,4$ are constant coefficients. The general
solution is a superposition of $\sinh x$ , $\cosh x$ , $\sin x$ and
$\cos x$. Considering  boundary condition in $x=0$ leads to the
result that $\cosh x$ and $ \cos x$ should not be considered in
physical solution. On the other hand, boundary condition in $x=a$
implies that $\sinh x$ should not be present in final solution since
in $x=a$ wave function should be vanishing. Therefore, to satisfy
boundary conditions only $\sin x$ is physically reasonable. It is
important to note that in the framework of quantum mechanical
perturbational approach, one recovers only $\sin x$ as leading order
solution, which approves the above statement. In other words,
considering the term proportional to $\beta$ in generalized
Schr\"{o}dinger equation as perturbation, one recovers $sin x$ as
the leading order wave function. Therefore, the solution of
generalized Schr\"{o}dinger equation with the prescribed boundary
conditions is
\begin{equation}
\psi(x) = A\sin
\Big[\frac{1}{2\hbar}\sqrt{\frac{1}{\beta}(\sqrt{1+16mE\beta
})-1}\Big]x.
\end{equation}
Defining
\begin{equation}
\frac{1}{2\hbar}\sqrt{\frac{1}{\beta}(\sqrt{1+16mE\beta })-1}=j
\end{equation}
boundary condition in $x=0$ leads to  $\sin ja=0=\sin n\pi$ where $
n= 1, 2, 3, ...$ . Therefore $ j=\frac{n\pi}{a}$. After
normalization one finds
\begin{equation}
\psi(x)=\sqrt{\frac{2}{a}}\sin\frac{n\pi x}{a}.
\end{equation}
Now the energy condition
\begin{equation}
\frac{1}{2\hbar}\sqrt{\frac{1}{\beta}(\sqrt{1+16mE\beta
})-1}=\frac{n\pi}{a},
\end{equation}
leads to
\begin{equation}
E_{n}=\frac{n^2\pi^2\hbar^2}{2ma^2}+\beta\frac{n^4\pi^4\hbar^4}{ma^4},
\end{equation}
up to first order in $\beta$. Note that in the framework of GUP up
to first order in $\beta$ there is no change in eigenstates of the
quantum particle in box but we have a shift of energy levels equal
to $ \beta\frac{n^4\pi^4\hbar^4}{ma^4}$. When $\beta\rightarrow 1$,
that is, in extreme quantum gravity limit, this energy shift plays
very important role. Figure 1 shows the corresponding situation. As
this figure shows increasing $\beta$ leads to increasing shift in
energy levels.

\section{Momentum Space Wave Function}
In ordinary quantum mechanics it is often useful to expand the
states $|\psi\rangle$ in the position eigenbasis $\{|x\rangle\}$ as
$\langle x|\psi\rangle$. As has been indicated, in GUP due to the
existence of minimal length scale which can not be probed, there are
no physical states to perform a position eigenbasis. Although there
is a one parameter family of $x$-eigenbases, related to the minimal
uncertainty in position, but these bases consists of no physical
states. Furthermore they could not even be approximated by physical
states of increasing localization. However we can still project
arbitrary states $|\phi\rangle$ on maximally localized states
$|\psi^{M.L.}_{x}\rangle$ to obtain the probability amplitude for
the particle being maximally localized around the position $x$[24].
In which follows we consider these maximally localized wave
functions.\\
The eigenvalue equation for momentum space wave function in ordinary
quantum mechanics is
\begin{equation}
P_{op}u_{p}(x)=pu_{p}(x)
\end{equation}
which $P_{op}$ stands for momentum operator with $p$ as its
eigenvalue and $u_{p}(x) $ as its  eigenfunction. Position space
representation of $P_{op}$ is as follows
\begin{equation}
P=\frac{\hbar}{i}\frac{\partial}{\partial x}.
\end{equation}
Therefore one finds
\begin{equation}
\frac{\hbar}{i}\frac{\partial u_{p}(x)}{\partial x}=pu_{p}(x)
\end{equation}
which has solution such as
\begin{equation}
u_{p}(x)=\frac{1}{\sqrt{2\pi\hbar}}e^{\frac{ipx}{\hbar}}.
\end{equation}
Now we consider gravitational induced uncertainty which leads to the
following generalized form for $P_{op}$
\begin{equation}
P_{op}=\frac{\hbar}{i}\frac{\partial}{\partial
x}[1+\beta(\frac{\hbar}{i}\frac{\partial}{\partial x})^2]
\end{equation}
up to first order in $\beta$. Eigenvalue equation in this case
should be modified as follows
\begin{equation}
\frac{\hbar}{i}\frac{\partial}{\partial
x}[1+\beta(\frac{\hbar}{i}\frac{\partial}{\partial
x})^2]u_{p}(x)=pu_{p}(x),
\end{equation}
which can be written as
\begin{equation}
\beta\hbar^3\frac{{\partial}^3 u_{p}(x)}{\partial
x^3}-\hbar\frac{\partial u_{p}(x)}{\partial x}+ipu_{p}(x)=0.
\end{equation}
This equation has a complicated solution
\begin{equation}
u_{p}(x)=C_{1}e^{-\frac{i\sqrt{3}(A-12\beta)+A+12\beta}{B}x}+C_{2}e^{-\frac{i\sqrt{3}(12\beta-A)+
A+12\beta}{B}x}+C_{3}e^{\frac{2(A+12\beta
)x}{B}},
\end{equation}
where $A$ and $B$ are complex quantities   $$
A=\Big(-12[9pi-\sqrt{-\frac{12+81p^2\beta}{\beta}}]\beta^{2}
\Big)^{2/3}$$ and $$ B=12\beta\hbar A^{1/2}$$ respectively. A
lengthy calculation gives the following result for momentum space
wave function up to first order of $\beta$.
\begin{equation}
u_{p}(x)=\Bigg(\frac{1-3\beta p^{2}}{2\pi\hbar}\Bigg)^{1/2}
\exp{\Bigg(\frac{i(p-\beta p^{3})x}{\hbar}\Bigg)}.
\end{equation}
It is important to note that this wave function reduces to (17) in
the limit of $\beta\rightarrow 0$ as a consequence of correspondence
principle.

\section{Summary and Conclusions}
Although position space representation fails to be valid in the case
of GUP due to existence of a minimal observable length, one can
consider maximally localized states to obtain the probability
amplitude for the particle being maximally localized around the
position $x$. In this paper, following such a viewpoint, we have
calculated eigenfunctions and eigenvalues of a spinless particle
confined in an one-dimensional potential box. We have shown that up
to first order in GUP parameter, there is no change in
eigenfunctions, but there is a shift to energy levels which has
considerable effect in the extreme quantum gravity
limit$\beta\rightarrow 1$. Following the maximally localized
picture, we have found momentum space wave function for a free
particle. Actually some care should be taken into account regarding
the notion of "free particle", since now due to gravitational effect
the particle is no longer free.

\begin{figure}[ht]
\begin{center}
\includegraphics{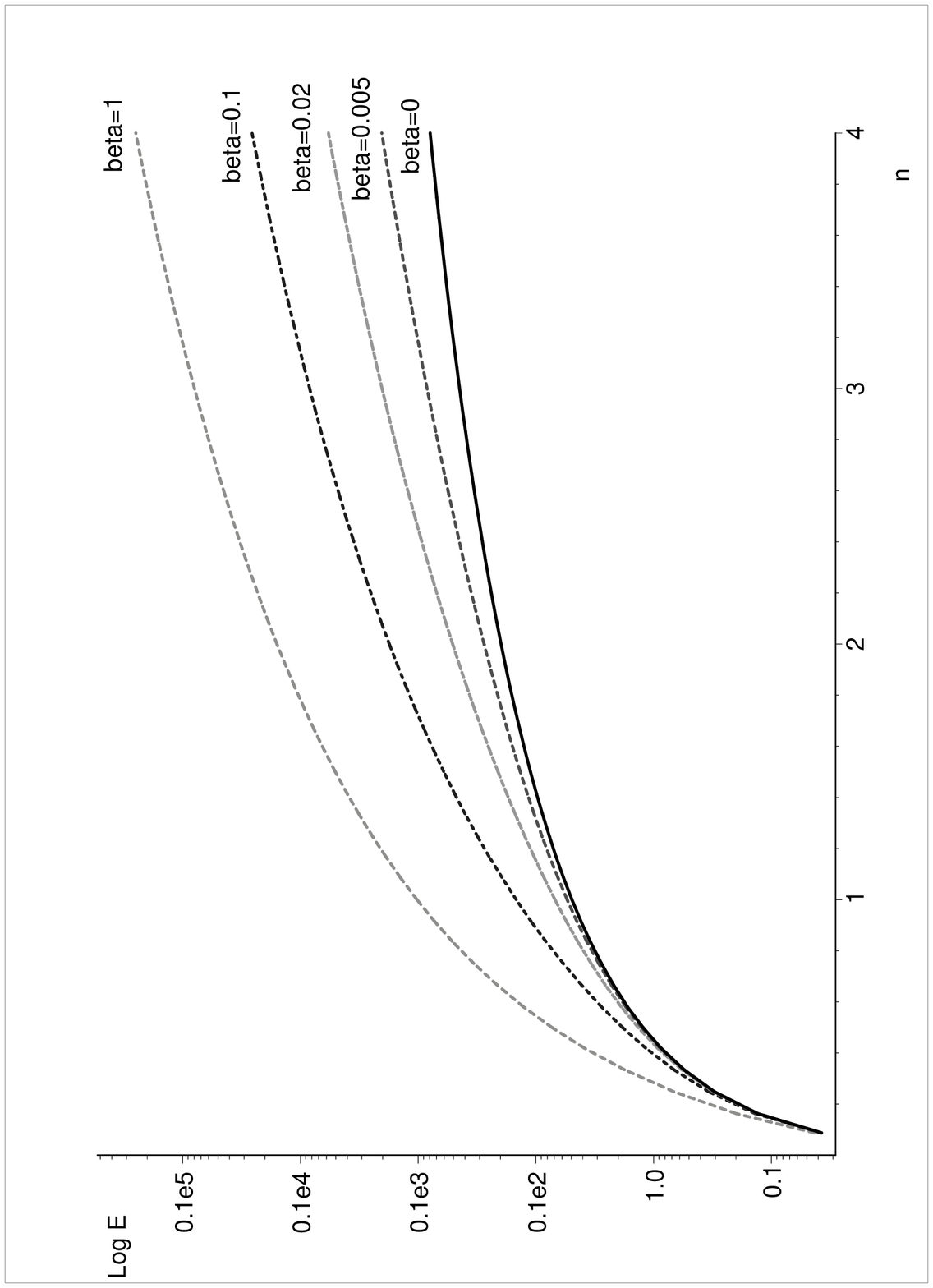}
\end{center}
\vspace{16 cm}
 \caption{\small {Energy Spectrum of a Particle in One-Dimensional Box in the Framework of GUP.
 The Effects of Variation in GUP Parameter($\beta$) is Highlighted.
 To Show the Effect of GUP, we have Considered $n$ as a Continuous Parameter. $\beta =0$ is the Usual Quantum Mechanical Regime.}}
 \label{fig:1}
\end{figure}

\end{document}